# On the Efficiency of 5(4) RK-Embedded Pairs with High Order Compact Scheme and Robin Boundary Condition for Options Valuation

**Chinonso Nwankwo[a,*], Weizhong Dai[b]**

[a] Modelling and Simulation, CEMSE, KAUST, Thuwal, Saudi Arabia

[b] Department of Mathematics and Statistics, Louisiana Tech University, Ruston LA 71272, USA

[*] Corresponding author, nonsonwankwo@gmail.com, chinonso.nwankwo@kaust.edu.sa

https://orcid.org/0000-0001-5526-1337

# On the Efficiency of 5(4) RK-Embedded Pairs with High Order Compact Scheme and Robin Boundary Condition for Options Valuation

## Abstract

When solving the American options with or without dividends, numerical methods often obtain lower convergence rates if further treatment is not implemented even using high-order schemes. In this article, we present a fast and explicit fourth-order compact scheme for solving the free boundary options. In particular, the early exercise features with the asset option and option sensitivity are computed based on a coupled of nonlinear PDEs with fixed boundaries for which a high order analytical approximation is obtained. Furthermore, we implement a new treatment at the left boundary by introducing a third-order Robin boundary condition. Rather than computing the optimal exercise boundary from the analytical approximation, we simply obtain it from the asset option based on the linear relationship at the left boundary. As such, a high order convergence rate can be achieved. We validate by examples that the improvement at the left boundary yields a fourth-order convergence rate without further implementation of mesh refinement, Rannacher time-stepping, and/or smoothing of the initial condition. Furthermore, we extensively compare, the performance of our present method with several 5(4) Runge-Kutta pairs and observe that Dormand and Prince and Bogacki and Shampine 5(4) pairs are faster and provide more accurate numerical solutions. Based on numerical results and comparison with other existing methods, we can validate that the present method is very fast and provides more accurate solutions with very coarse grids.

## 1. Introduction

Suppose $P(S,\tau)$ and $f_b(\tau)$ and $E$ are the put option price, optimal exercise boundary, and strike price, respectively, and $\tau = T - t$. Then, $P(S,\tau)$ satisfies the free boundary value problem:

$$\frac{\partial P(S,\tau)}{\partial \tau} - \frac{1}{2}\sigma^2 S^2 \frac{\partial^2 P(S,\tau)}{\partial S^2} - (\mu - \delta) S \frac{\partial P(S,\tau)}{\partial S} + (\mu - \varrho) P(S,\tau) = \zeta(S,\tau), \qquad S > f_b(\tau); \qquad (1a)$$

$$P(S,\tau) = E - S, \qquad S < f_b(\tau). \qquad (1b)$$

Here, $\zeta$ and $\varrho$ could represent models which include the jump-diffusion model, regime-switching model, one-dimensional FX options, or American options. If $\mu \in [r, r_d]$, $\delta \in [r_f, D, 0]$, $\zeta = 0$, and $\varrho = 0$, we have a model suitable for solving American and FX options. In this case, $r_d$ and $r_f$ are the domestic and foreign interest rates, respectively, for one-dimensional FX options. $r$ and $D$ are the interest rate and the dividend, respectively, for non-dividend and dividend American options. If we consider the latter, (1) reduces to

$$\frac{\partial P(S,\tau)}{\partial \tau} - \frac{1}{2}\sigma^2 S^2 \frac{\partial^2 P(S,\tau)}{\partial S^2} - (r-D)S\frac{\partial P(S,\tau)}{\partial S} + rP(S,\tau) = 0, \qquad S > f_b(\tau); \qquad (2a)$$

$$P(S,\tau) = E - S, \qquad S < f_b(\tau). \qquad (2b)$$

The initial and boundary conditions are given as

$$P(S,0) = max(E-S,0), \qquad f_b(0) = E; \qquad (2c)$$

$$P(f_b,\tau) = E - f_b(\tau), \quad P(0,\tau) = E, \quad P(\infty,\tau) = 0, \quad \frac{\partial}{\partial S}P(f_b,\tau) = -1. \qquad (2d)$$

After the first algorithm for solving the American option pricing model was proposed by Brennan and Schwartz [5], other several methods have further been introduced. For the American options with or without dividends, mostly, the first-order and second-order explicit and implicit schemes have been implemented [10,11,20,30,37,42,44]. This is because the first and second-order convergence rates can easily be obtained using these schemes. Few authors have proposed high-order numerical methods for solving American style options with or without dividends [1,7,9,18,38,46]. Some of these authors that implemented high order numerical scheme either did not provide the result of the convergence rate or obtained a numerical convergence rate that is either not in good agreement with the theoretical convergence rate [18,38,46] or unstable. For instance, even with an efficient fourth-order compact scheme, only the second-order convergence rate should be expected. Certain improvements have been proposed for recovering high order convergence rate which includes grid refinement and stretching [31], smoothing of the initial condition [9,16,33], repeated Richardson extrapolation [1], etc. Some of these improvements are either difficult to implement or require an additional computational cost. This increases the overall computational burden for approximating American-style options.

Here, to reduce computational cost substantially and recover the fourth-order convergence rate with a fourth-order compact scheme, we first implement a special treatment from the information on the left boundary. We then propose a simple, efficient, and fast fourth-order compact scheme with Robin boundary condition and RK-embedded 5(4) pairs for approximating (2). It is well known that in the location of large variation and sufficient smoothness, smaller and larger time steps, respectively, might be required to achieve more accurate numerical approximations. Several embedded Runge-Kutta pairs have been developed and implemented to address this issue in the literature

[3,4,6,13,15,21,23,27,32,34,35,36,39,40,41]. It is worth noting that the performance of some of these embedded pairs is model-dependent. Even though most of the option pricing models exhibit some form of discontinuity and variation, however, these embedded pairs have not attracted much attention in options valuation. It is in our interest to investigate the performance of our present method with several 5(4) RK-embedded pairs and validate the ones more suitable for solving non-dividend and dividend American options.

The rest of the paper is organized as follows. In section 2, we present a semi-analytical approach for computing the time-dependent coefficient that introduces the nonlinearity in the transformed model. In section 3, we present the high-order numerical scheme for the approximation of the optimal exercise boundary simultaneously with the asset option and option sensitivity. In section 4, we discuss our numerical experiment and compare the performance of our proposed method with several 5(4) RK-embedded pairs. Furthermore, we compare our numerical results with the well-known existing methods and conclude the paper in section 5.

## 2. Semi-Analytical Formulation with Time-Dependent Coefficient

Considering the free boundary problem for pricing options, the early exercise feature needs to be computed simultaneously with the options. To achieve this possibility, we first fix the free boundary by introducing the transformed relationship [10,14,42] below to (2)

$$x = \ln S - \ln s_f(\tau), \qquad U(x,\tau) = P(S,\tau). \tag{3}$$

With this transformation, we obtained a coupled nonlinear PDE equation for the asset option as follows:

$$U_\tau(S,\tau) - \frac{1}{2}\sigma^2 U_{xx}(S,\tau) - \omega(\tau)V(S,\tau) + rU(S,\tau) = 0, \qquad x > 0; \tag{4a}$$

$$U(x,\tau) = K - S = K - s_f(\tau)e^x, \qquad x < 0. \tag{4b}$$

Taking the derivative of (4a) with respect to $x$ and denoting $V = U_x$, we obtain a coupled nonlinear PDE equation for delta option as follows:

$$V_\tau(S,\tau) - \frac{1}{2}\sigma^2 V_{xx}(S,\tau) - \omega(\tau)\frac{1}{2}\sigma^2 U_{xx} + rV(S,\tau) = 0, \qquad x > 0; \tag{4c}$$

$$V(x,\tau) = -S = -s_f(\tau)e^x, \qquad x < 0. \tag{4d}$$

The initial and boundary conditions for the asset option and option sensitivity are given as follows:

$$U(x,0) = \max(K - Ke^x, 0) = 0, \qquad x \geq 0, \qquad s_f(0) = K; \tag{4e}$$

$$U(0,\tau) = K - s_f(\tau), \qquad U(\infty,\tau) = 0, \tag{4f}$$

$$V(0,\tau) = -s_f(\tau), \qquad V(\infty,\tau) = 0. \tag{4g}$$

The non-linearity in the transformed coupled system of PDEs is due to the coefficient of the convective term

$$\omega(\tau) = (r - D) + \frac{f_b'(\tau)}{f_b(\tau)} - \frac{\sigma^2}{2}, \tag{5}$$

which is time-dependent and involves the derivative of the optimal exercise boundary. By implementing a front-fixing approach, we focus on the space domain $[0,\infty)$ with the fixed boundary representing the left boundary point. It is important to mention that with this approach, the non-smoothness in the second derivative of the asset option is no longer visible (see the work of Ballestra [1]). However, another source of non-smoothness arises due to the time-dependent coefficient in (5). This is because $f_b(\tau)$ is not differentiable when $\tau = 0$ (see Ballestra [1], Chen and Chadam [8], Mallier [28]). Chen and Chadam [8] explained that the non-differentiability of $f_b(\tau)$ when $\tau = 0$ is due to the non-smoothness of the maximum function at the payoff. Ballestra [1] further mentioned that if the source of this non-smoothness is efficiently handled, then a high order convergent rate can be expected with the front-fixing approach.

To handle this source of non-smoothness, Ballestra [1] implemented a time variable transformation. Here, we derive a high order analytical formulation for obtaining a precise approximation of (5) without time discretization and then approximate (3) with high order compact finite difference scheme. To this end, we first introduce a transformation with intermediate function (that has Lipschitz character near the left boundary) as follows [24,25]:

$$L(x,\tau) = \sqrt{U(x,\tau) - E + e^x f_b(\tau)}, \tag{6a}$$

with

$$L(x,\tau)\begin{cases} = 0, & x \in \left[\ln s_f(\infty) - \ln s_f(0)\right], \\ > 0, & x \in (0,\infty). \end{cases} \tag{6b}$$

Here, we consider the derivative of the intermediate function up to the third-order derivative as follows:

$$U(x,\tau) = L^2(x,\tau) + E - e^x f_b(\tau), \qquad U(0,\tau) = E - f_b(\tau), \tag{7a}$$

$$U_x = 2LL' - e^x f_b(\tau), \qquad U_x(0,\tau) = -f_b(\tau). \tag{7b}$$

Following the work of Goodman and Ostrov [17], we differentiate $U(x,\tau)$ with respect to $\tau$ when $x = 0$ as follows:

$$U_x \ln 1 + U_\tau = -f_b'(\tau), \qquad U_\tau = -f_b'(\tau). \tag{7c}$$

Furthermore,

$$U_{xx} = 2(L')^2 + 2LL'' - e^x f_b(\tau), \qquad U_{xx}(0,\tau) = 2\big(L'(0,\tau)\big)^2 - f_b(\tau). \tag{7d}$$

Substituting (6) in (3a), we obtain

$$L'(0,\tau) = \frac{\sqrt{rE - Df_b(\tau)}}{\sigma}. \tag{7e}$$

The analysis of (7e) is similar to the one presented in the work of Kim et al. [24]. We would further love to point out that when $D \neq 0,\ r \geq Df_b(\tau)/E$ is required to ensure a real value of $L'(0,\tau)$. Hence at the payoff when $f_b(0) = E$, $r \geq D$ will ensure a real value of $L'(0,\tau)$. If $D = 0$, such a condition will not be imposed.

Next, we compute the derivative of $U(x,\tau)$ in (3b) as follows:

$$U_{xxx} = 6L'L'' + 2LL''' - e^x f_b(\tau), \qquad U_{xxx}(0,\tau) = 6L'L'' - f_b(\tau); \tag{8a}$$

$$U_{xx} \ln 1 + U_{x\tau} = -f_b'(\tau), \qquad U_{x\tau}(0,\tau) = -f_b'(\tau). \tag{8b}$$

Substituting (8) in (3b), we obtain

$$L''(0,\tau) = -\frac{2L'(0,\tau)}{3\sigma^2}\xi_\tau - \frac{2L'(0,\tau)}{3\sigma^2}\kappa - \frac{Df_b(\tau)}{3\sigma^2 L'(0,\tau)}. \tag{9}$$

Here, $\xi_\tau = f_b'(\tau)/f_b(\tau)$ and $\kappa = r - D - \sigma^2/2$. Finally, we further take the derivative of (3b) with respect to $x$ and obtain:

$$U_{xx\tau}(S,\tau) - \frac{1}{2}\sigma^2 U_{xxxx}(S,\tau) - \omega(\tau)U_{xxx} + rU_{xx} = 0. \tag{10}$$

Computing the partial derivatives in (8), we obtain as follows:

$$U_{xxxx} = 6(L'')^2 + 8L'L''' - e^x f_b(\tau), \qquad U_{xxxx}(0,\tau) = 2(L\,')^2 - f_b(\tau); \tag{11a}$$

$$U_{xxx} \ln 1 + U_{xx\tau} = -\frac{2Df_b'(\tau)}{\sigma^2} - f_b'(\tau), \qquad U_{xx\tau}(0,\tau) = -\frac{2Df_b'(\tau)}{\sigma^2} - f_b'(\tau). \tag{11b}$$

Substituting (9) in (8), we obtain

$$L'''(0,\tau) = \frac{2L'(0,\tau)}{3\sigma^4}\xi_\tau^2 + \left[\frac{4L'(0,\tau)\kappa}{3\sigma^4} - \frac{2Df_b(\tau)}{3\sigma^2 L'(0,\tau)}\right]\xi_\tau + \frac{2L'(0,\tau)\kappa^2}{3\sigma^4} + \frac{Df_b(\tau)\kappa}{6\sigma^4 L'(0,\tau)} - \frac{\left(Df_b(\tau)\right)^2}{12\sigma^4[L'(0,\tau)]^3}$$
$$+ \frac{rL'(0,\tau)}{2\sigma^2} - \frac{Df_b(\tau)}{4\sigma^2 L'(0,\tau)}. \tag{12}$$

It is worth mentioning that $\xi_\tau$ is involved in the interior discretization of the system of PDEs. Moreover, we intend to discretize in space using a fourth-order compact finite difference scheme. Hence, we need to derive the analytical approximation of $\xi_\tau$ up to the fourth-order accuracy in space. To this end, we introduce an extrapolated Taylor series expansion of $L(\tilde{x},\tau)$ at $x = 0$ up to the seventh order accuracy as follows:

$$\alpha_0 L(0,\tau) + \alpha_1 L(\tilde{x},\tau) + \alpha_2 L(2\tilde{x},\tau) + \alpha_3 L(3\tilde{x},\tau) + \alpha_4 L(4\tilde{x},\tau)$$
$$= \gamma_0 \tilde{x} L'(0,\tau) + \gamma_1 \tilde{x}^2 L''(0,\tau) + \gamma_2 \tilde{x}^3 L'''(0,\tau) + O(\bar{x}^7). \tag{13a}$$

Here,

$$\alpha_0 = -\frac{3445}{27}, \qquad \alpha_1 = 256, \qquad \alpha_2 = -48, \qquad \alpha_3 = \frac{256}{27}, \qquad \alpha_4 = -1; \tag{13b}$$

$$\gamma_0 = \frac{4980}{27}, \qquad \gamma_1 = \frac{600}{9}, \qquad \gamma_2 = \frac{32}{3}. \tag{13c}$$

With Taylor series expansion and method of extrapolation, it is straightforward to obtain the coefficients in (13b) and (13c). Hence, for brevity, we skip its detailed derivation. Here, $\tilde{x} \ll x$. Substituting (7), (9), and (12) into (13a), we obtain a high order analytical approximation of $\xi_\tau$ in quadratic form as follows:

$$g_2(\xi_\tau)^2 + g_1\xi_\tau + g_0 = 0, \tag{14a}$$

Because $f_b' < 0$ and $f_b > 0$, it implies that $\xi_\tau < 0$. For real value solutions of (14a), we have it that

$$\xi_\tau = \frac{-g_1 - \sqrt{g_1^2 - 4g_2 g_{0,4}}}{2g_2}, \quad \omega(\tau) = \xi_\tau + \kappa. \tag{14b}$$

Here,

$$g_2 = \frac{2L'(0,\tau)\gamma_2\tilde{x}^3}{3\sigma^4}, \tag{14c}$$

$$g_1 = \left[\frac{4L'(0,\tau)\kappa}{3\sigma^4} - \frac{2Df_b(\tau)}{3\sigma^2 L'(0,\tau)}\right]\gamma_2\tilde{x}^3 - \frac{2L'(0,\tau)\gamma_1\tilde{x}^2}{3\sigma^2}, \tag{14d}$$

$$g_{0,4} = \left[\frac{2L'(0,\tau)\kappa^2}{3\sigma^4} + \frac{Df_b(\tau)\kappa}{6\sigma^4 L'(0,\tau)} - \frac{\left(Df_b(\tau)\right)^2}{12\sigma^4[L'(0,\tau)]^3} + \frac{rL'(0,\tau)}{2\sigma^2} - \frac{Df_b(\tau)}{4\sigma^2 L'(0,\tau)}\right]\gamma_2\tilde{x}^3$$

$$-\left[\frac{2L'(0,\tau)}{3\sigma^2}\kappa + \frac{Df_b(\tau)}{3\sigma^2 L'(0,\tau)}\right]\gamma_1\tilde{x}^2 + \gamma_0\tilde{x}L'(0,\tau) - M_4(\tilde{x},\tau), \qquad (14d)$$

$$M_4(\tilde{x},\tau) = \alpha_1 L(\tilde{x},\tau) + \alpha_2 L(2\tilde{x},\tau) + \alpha_3 L(3\tilde{x},\tau) + \alpha_4 L(4\tilde{x},\tau). \qquad (14e)$$

## 3. Numerical Discretization

To construct our numerical scheme, we first define our computational domain. Because the asset and delta option options vanish rapidly as the value of $x$ increases, the infinite space domain $[0,\infty) \times [0,T]$ is replaced with a bounded domain $[0,x_M] \times [0,T]$ and the right boundary is given as

$$U(x_M,\tau) = V(x_M,\tau) = 0. \qquad (15)$$

Furthermore, we consider a uniform grid size as follows:

$$x_i = ih, \qquad h = \frac{x_M}{M}, \qquad i \in [0,M], \qquad (16)$$

$M$ is the total number of grid points. The solutions of the asset option, option sensitivity, and the optimal exercise boundary are given as $u_i^n$, $v_i^n$, and $f_b^n$, respectively.

### 3.1. Numerical Schemes

In our present method, the optimal exercise boundary is computed from the asset option based on their linear relationship as follows:

$$U(0,\tau) = E - f_b(\tau). \qquad (17)$$

To this end, we introduce a boundary condition for computing the asset option when $x = 0$ by considering the following lemma.

**Lemma**. Assume $f(x) \in C^5[x_0,x_1]$, then it holds

$$\frac{7}{24}f''(x_0) + \frac{6}{24}f''(x_1) - \frac{1}{24}f''(x_2) = \frac{1}{h^2}[f(x_1) - f(x_0)] - \frac{1}{h}f'(x_0) + O(h^3). \qquad (18)$$

**Proof.** For (18a), we first introduce a well-known third-order boundary condition [43] as follows:

$$\frac{10}{24}f''(x_0) + \frac{2}{24}f''(x_1) = \frac{1}{h^2}[f(x_1) - f(x_0)] - \frac{1}{h}f'(x_0) - \frac{h}{12}f^{(3)}(x_0) + O(h^3). \qquad (19a)$$

Next, we approximate the third-order derivative as follows:

$$\frac{h}{12}f'''(x_0) = -\frac{1}{8}f''(x_0) + \frac{1}{6}f''(x_1) - \frac{1}{24}f''(x_2) + O(h^3). \tag{19b}$$

Substituting (19b) into (19a), we then complete the proof for (18). For the interior nodes, we consider a fourth-order compact finite difference scheme as follows:

$$f''(x_{i-1}) + 10f''(x_i) + f''(x_{i+1}) = \frac{12}{h^2}[f(x_{i-1}) - 2f(x_i) + f(x_{i+1})] + O(h^4). \tag{20}$$

By introducing a Robin boundary condition based on the relationship between the asset option, option sensitivity, and optimal exercise boundary as follows:

$$V(0,\tau) - U(0,\tau) = -K, \tag{21a}$$

where $V = U_x$, the matrix-vector form for the asset option is as follows:

$$A = \frac{12}{h^2}\begin{bmatrix} -2a & 2 & 0 & \cdots & & & & 0 \\ 1 & -2 & 1 & & & & & \vdots \\ & 1 & -2 & 1 & & & & \\ & & 1 & -2 & 1 & & & \\ 0 & & & \ddots & \ddots & \ddots & & 0 \\ & & & & 1 & -2 & 1 & \\ \vdots & & & & & 1 & -2 & 1 \\ 0 & & & & \cdots & 0 & 1 & -2 \end{bmatrix}_{M\times M},$$

$$B = \begin{bmatrix} 7 & 6 & -1 & 0 & 0 & \cdots & 0 \\ 1 & 10 & 1 & & & & \vdots \\ & 1 & 10 & 1 & & & \\ & & 1 & 10 & 1 & & \\ 0 & & & \ddots & \ddots & \ddots & 0 \\ \vdots & & & & 1 & 10 & 1 \\ 0 & \cdots & 0 & 0 & 0 & 0 & 10 & 1 \end{bmatrix}_{M\times M}, \quad \boldsymbol{f}_u = \frac{24}{h}\begin{bmatrix} K \\ 0 \\ 0 \\ \vdots \\ 0 \\ 0 \end{bmatrix}_{M\times 1}; \tag{21b}$$

$$\boldsymbol{u}'' = B^{-1}(A\boldsymbol{u}^n + \boldsymbol{f}_u^n), \tag{21c}$$

$$\frac{\partial \boldsymbol{u}}{\partial \tau} = \boldsymbol{v}(\boldsymbol{u},\boldsymbol{v}) = \frac{\sigma^2}{2}B^{-1}(A\boldsymbol{u}^n + \boldsymbol{f}_u^n) + \omega(\tau)\boldsymbol{v} - r\boldsymbol{u}. \tag{21d}$$

Here, $a = 1 + h > 1$. It is worth mentioning that the third-order Robin boundary preserves diagonal dominance in both matrices. The boundary-value of the delta option can easily be computed from the optimal exercise boundary as follows:

$$V(0,\tau) = -f_b(\tau). \tag{22}$$

For the delta option, we discretize the interior nodes with (20). For $i = 1$ and $i = M - 1$, we employ a discretization as follows [45,2]:

$$14f''(x_1) - 5f''(x_2) + 4f''(x_3) - f''(x_4) = \frac{12}{h^2}[f(x_0) - 2f(x_1) + f(x_2)] + O(h^4), \qquad (23a)$$

$$14f''(x_{M-1}) - 5f''(x_{M-2}) + 4f''(x_{M-3}) - f''(x_{M-4}) = \frac{12}{h^2}[f(x_{M-2}) - 2f(x_{M-1}) + f(x_M)] + O(h^4). \qquad (23b)$$

The matrix-vector form for the delta option is as follows:

$$F = \frac{12}{h^2}\begin{bmatrix} -2 & 1 & 0 & \cdots & & & & 0 \\ 1 & -2 & 1 & & & & & \vdots \\ & 1 & -2 & 1 & & & & \\ & & 1 & -2 & 1 & & & \\ 0 & & & \ddots & \ddots & \ddots & & 0 \\ & & & & 1 & -2 & 1 & \\ \vdots & & & & & 1 & -2 & 1 \\ 0 & & & & \cdots & 0 & 1 & -2 \end{bmatrix}_{M-1\times M-1},$$

$$G = \begin{bmatrix} 14 & -5 & 4 & -1 & 0 & \cdots & 0 \\ 1 & 10 & 1 & & & & \vdots \\ & 1 & 10 & 1 & & & \\ & & 1 & 10 & 1 & & \\ 0 & & & \ddots & \ddots & \ddots & 0 \\ \vdots & & & & 1 & 10 & 1 \\ 0 & \cdots & 0 & -1 & 4 & -5 & 14 \end{bmatrix}_{M-1\times M-1}', \quad \boldsymbol{f}_v = \frac{12}{h^2}\begin{bmatrix} v_0 \\ 0 \\ 0 \\ \vdots \\ 0 \\ v_M = 0 \end{bmatrix}_{M-1\times 1}; \qquad (24a)$$

$$\boldsymbol{v}'' = G^{-1}(F\boldsymbol{v} + \boldsymbol{f}_v), \qquad (24b)$$

$$\frac{\partial \boldsymbol{v}}{\partial \tau} = v(\boldsymbol{u}, \boldsymbol{v}) = \frac{\sigma^2}{2} G^{-1}(\boldsymbol{F}\boldsymbol{v} + \boldsymbol{f}_v) + \omega(\tau) G^{-1}(F\boldsymbol{u} + \boldsymbol{f}_u) - r\boldsymbol{v}. \qquad (24c)$$

For the time integration, we compare the performance and accuracy of several 5(4) embedded Runge-Kutta methods. Specifically, we consider Dormand and Prince 5(4) [13], Cash and Karp 5(4) [6], Bogacki and Shampine 5(4) [4], Simos and Tsitouras 5(4) [36], and Papakostas and Papageorgiou 5(4) [32] embedded pairs. We only describe the effective implementation of Dormand and Prince 5(4) and refer the readers to the work of Bogacki and Shampine [4], Cash and Karp [6] Simos and Tsitouras [36] Papakostas and Papageorgiou [32], and Ketcheson et al. [23] on how to obtain the coefficient entries of the other Runge-Kutta pairs which we considered in this work. To this end, we first present the semi-discretized coupled nonlinear equations in (21) and (24) as follows:

$$\frac{\partial \boldsymbol{u}^n}{\partial \tau} = \boldsymbol{v}^n = \frac{\sigma^2}{2} B^{-1}(A\boldsymbol{u}^n + \boldsymbol{f}_u^n) + \omega^n(\tau)\boldsymbol{v}^n - r\boldsymbol{u}^n, \qquad (25a)$$

$$\frac{\partial \boldsymbol{v}^n}{\partial \tau} = \boldsymbol{v}^n = \frac{\sigma^2}{2} G^{-1}(G\boldsymbol{v}^n + \boldsymbol{f}_v^n) + \omega^n(\tau) G^{-1}(F\boldsymbol{u}^n + \boldsymbol{f}_u^n) - r\boldsymbol{v}^n. \qquad (25b)$$

The error for the embedded pairs is estimated as

$$e_u = \|\widetilde{\boldsymbol{u}}^{n+1} - \boldsymbol{u}^{n+1}\|_\infty, \quad e_v = \|\widetilde{\boldsymbol{v}}^{n+1} - \boldsymbol{v}^{n+1}\|_\infty. \tag{25c}$$

Here, $\boldsymbol{u}^{n+1}$ and $\boldsymbol{v}^{n+1}$ are the numerical approximations of the asset option and option sensitivity with the fifth-order accuracy in time and the fourth-order accuracy in space, respectively. We would love to point out that we consider both the error estimate of the asset option and option sensitivity in this work. $\widetilde{\boldsymbol{u}}^{n+1}, \boldsymbol{u}^{n+1}, f_b^{n+1}, \widetilde{\boldsymbol{v}}^{n+1}$, and $\boldsymbol{v}^{n+1}$ are computed simultaneously with Dormand and Prince 5(4) coefficients as follows:

$$\boldsymbol{u}^{n(1)} = \boldsymbol{u}^n, \qquad \boldsymbol{v}^{n(1)} = \boldsymbol{v}^n, \qquad f_b^{n(1)} = f_b^n; \tag{26a}$$

$$\xi_{n(1)} = \frac{-g_1 - \sqrt{g_1^2 - 4g_2 g_{0,4,n(1)}}}{2g_2}, \qquad \omega^{n(1)} = \xi_{n(1)} + \kappa; \tag{26b}$$

$$\boldsymbol{R}_u^1 = v\big(\boldsymbol{u}^{n(1)}, \boldsymbol{v}^{n(1)}\big)k, \qquad \boldsymbol{R}_v^1 = v\big(\boldsymbol{u}^{n(1)}, \boldsymbol{v}^{n(1)}\big)k; \tag{26c}$$

$$\boldsymbol{u}^{n(2)} = \boldsymbol{u}^n + \frac{k}{5}\boldsymbol{R}_u^1, \qquad f_b^{n(2)} = K - \boldsymbol{u}_1^{n(2)}, \qquad \boldsymbol{v}_1^{n(2)} = -f_b^{n(2)}, \qquad \boldsymbol{v}^{n(2)} = \boldsymbol{v}^n + \frac{k}{5}\boldsymbol{R}_v^1; \tag{26d}$$

$$\xi_{n(2)} = \frac{-g_1 - \sqrt{g_1^2 - 4g_2 g_{0,4,n(2)}}}{2g_2}, \qquad \omega^{n(2)} = \xi_{n(2)} + \kappa; \tag{26e}$$

$$\boldsymbol{R}_u^2 = v\big(\boldsymbol{u}^{n(2)}, \boldsymbol{v}^{n(2)}\big)k, \qquad \boldsymbol{R}_w^2 = v\big(\boldsymbol{u}^{n(2)}, \boldsymbol{v}^{n(2)}\big)k; \tag{26f}$$

$$\boldsymbol{u}^{n(3)} = \boldsymbol{u}^n + \frac{3k}{40}\boldsymbol{R}_u^1 + \frac{9k}{40}\boldsymbol{R}_u^2, \qquad f_b^{n(3)} = K - \boldsymbol{u}_1^{n(3)}; \tag{26g}$$

$$\boldsymbol{v}_1^{n(3)} = -f_b^{n(3)}, \qquad \boldsymbol{v}^{n(3)} = \boldsymbol{v}^n + \frac{3k}{40}\boldsymbol{R}_v^1 + \frac{9k}{40}\boldsymbol{R}_v^2; \tag{26h}$$

$$\xi_{n(3)} = \frac{-g_1 - \sqrt{g_1^2 - 4g_2 g_{0,4,n(3)}}}{2g_2}, \qquad \omega^{n(3)} = \xi_{n(3)} + \kappa; \tag{26i}$$

$$\boldsymbol{R}_u^3 = v\big(\boldsymbol{u}^{n(3)}, \boldsymbol{v}^{n(3)}\big)k, \qquad \boldsymbol{R}_v^3 = v\big(\boldsymbol{u}^{n(3)}, \boldsymbol{v}^{n(3)}\big)k; \tag{26j}$$

$$\boldsymbol{u}^{n(4)} = \boldsymbol{u}^n + \frac{44k}{45}\boldsymbol{R}_u^1 - \frac{56k}{15}\boldsymbol{R}_u^2 + \frac{32k}{9}\boldsymbol{R}_u^3, \qquad f_b^{n(4)} = K - \boldsymbol{u}_1^{n(4)}; \tag{26k}$$

$$\boldsymbol{v}_1^{n(4)} = -f_b^{n(4)}, \qquad \boldsymbol{v}^{n(4)} = \boldsymbol{v}^n + \frac{44k}{45}\boldsymbol{R}_v^1 - \frac{56k}{15}\boldsymbol{R}_v^2 + \frac{32k}{9}\boldsymbol{R}_v^3; \tag{26l}$$

$$\xi_{n(4)} = \frac{-g_1 - \sqrt{g_1^2 - 4g_2 g_{0,4,n(4)}}}{2g_2}, \qquad \omega^{n(4)} = \xi_{n(4)} + \kappa; \tag{26m}$$

$$\boldsymbol{R}_u^4 = v\big(\boldsymbol{u}^{n(4)}, \boldsymbol{v}^{n(4)}\big)k, \qquad \boldsymbol{R}_v^3 = \nu\big(\boldsymbol{u}^{n(4)}, \boldsymbol{v}^{n(4)}\big)k; \tag{26n}$$

$$\boldsymbol{u}^{n(5)} = \boldsymbol{u}^n + \frac{19732k}{6561}\boldsymbol{R}_u^1 - \frac{25360k}{2187}\boldsymbol{R}_u^2 + \frac{64448k}{6561}\boldsymbol{R}_u^3 - \frac{212k}{729}\boldsymbol{R}_u^4, \qquad f_b^{n(5)} = K - \boldsymbol{u}_1^{n(5)}; \tag{26o}$$

$$\boldsymbol{v}_1^{n(5)} = -f_b^{n(5)}, \qquad \boldsymbol{v}^{n(5)} = \boldsymbol{v}^n + \frac{19732k}{6561}\boldsymbol{R}_v^1 - \frac{25360k}{2187}\boldsymbol{R}_v^2 + \frac{64448k}{6561}\boldsymbol{R}_v^3 - \frac{212k}{729}\boldsymbol{R}_v^4; \tag{26p}$$

$$\xi_{n(5)} = \frac{-g_1 - \sqrt{g_1^2 - 4g_2 g_{0,4,n(5)}}}{2g_2}, \qquad \omega^{n(5)} = \xi_{n(5)} + \kappa; \tag{26q}$$

$$\boldsymbol{R}_u^5 = v\big(\boldsymbol{u}^{n(5)}, \boldsymbol{v}^{n(5)}\big)k, \qquad \boldsymbol{R}_v^5 = \nu\big(\boldsymbol{u}^{n(5)}, \boldsymbol{v}^{n(5)}\big)k; \tag{26r}$$

$$\boldsymbol{u}^{n(6)} = \boldsymbol{u}^n + \frac{9017k}{3168}\boldsymbol{R}_u^1 - \frac{355k}{33}\boldsymbol{R}_u^2 + \frac{46732k}{5247}\boldsymbol{R}_u^3 + \frac{49k}{176}\boldsymbol{R}_u^4 - \frac{5103k}{18656}\boldsymbol{R}_u^5, \qquad f_b^{n(6)} = K - \boldsymbol{u}_1^{n(6)}; \tag{26s}$$

$$\boldsymbol{v}_1^{n(6)} = -f_b^{n(6)}, \qquad \boldsymbol{v}^{n(6)} = \boldsymbol{v}^n + \frac{9017k}{3168}\boldsymbol{R}_v^1 - \frac{355k}{33}\boldsymbol{R}_v^2 + \frac{46732k}{5247}\boldsymbol{R}_v^3 + \frac{49k}{176}\boldsymbol{R}_v^4 - \frac{5103k}{18656}\boldsymbol{R}_v^5; \tag{26t}$$

$$\xi_{n(6)} = \frac{-g_1 - \sqrt{g_1^2 - 4g_2 g_{0,4,n(6)}}}{2g_2}, \qquad \omega^{n(6)} = \xi_{n(6)} + \kappa; \tag{26u}$$

$$\boldsymbol{R}_u^6 = v\big(\boldsymbol{u}^{n(6)}, \boldsymbol{v}^{n(6)}\big)k, \qquad \boldsymbol{R}_v^6 = \nu\big(\boldsymbol{u}^{n(6)}, \boldsymbol{v}^{n(6)}\big)k; \tag{26v}$$

$$\boldsymbol{u}^{n(7)} = \boldsymbol{u}^n + \frac{9017k}{3168}\boldsymbol{R}_u^1 + \frac{500k}{1113}\boldsymbol{R}_u^3 + \frac{125k}{192}\boldsymbol{R}_u^4 - \frac{2187k}{6784}\boldsymbol{R}_u^5 + \frac{11k}{84}\boldsymbol{R}_u^6, \qquad f_b^{n(7)} = K - \boldsymbol{u}_1^{n(7)}; \tag{26w}$$

$$\boldsymbol{v}_1^{n(7)} = -f_b^{n(7)}, \qquad \boldsymbol{v}^{n(7)} = \boldsymbol{v}^n + \frac{9017k}{3168}\boldsymbol{R}_v^1 + \frac{500k}{1113}\boldsymbol{R}_v^3 + \frac{125k}{192}\boldsymbol{R}_v^4 - \frac{2187k}{6784}\boldsymbol{R}_v^5 + \frac{11k}{84}\boldsymbol{R}_v^6; \tag{26x}$$

$$\xi_{n(7)} = \frac{-g_1 - \sqrt{g_1^2 - 4g_2 g_{0,4,n(7)}}}{2g_2}, \qquad \omega^{n(7)} = \xi_{(7)} + \kappa; \tag{26y}$$

$$\boldsymbol{R}_u^7 = v\big(\boldsymbol{u}^{n(7)}, \boldsymbol{v}^{n(7)}\big)k, \qquad \boldsymbol{R}_v^7 = \nu\big(\boldsymbol{u}^{n(7)}, \boldsymbol{v}^{n(7)}\big)k. \tag{26z}$$

The approximation with fifth-order Runge Kutta method given as

$$\boldsymbol{u}^{n+1} = \boldsymbol{u}^{n(7)}, \qquad f_b^{n+1} = f_b^{n(7)}, \qquad \boldsymbol{v}^{n+1} = \boldsymbol{v}^{n(7)}, \qquad \xi_{n+1} = \xi_{n(7)}, \qquad \omega^{n+1} = \omega^{n(7)}. \tag{27a}$$

represents the numerical solutions while the fourth-order Runge Kutta method given as

$$\tilde{\boldsymbol{u}}^{n+1} = \boldsymbol{u}^n + \frac{5179k}{57600}\boldsymbol{R}_u^1 + \frac{7571k}{16695}\boldsymbol{R}_u^3 + \frac{393k}{640}\boldsymbol{R}_u^4 - \frac{92097k}{339200}\boldsymbol{R}_u^5 + \frac{187k}{2100}\boldsymbol{R}_u^6 + \frac{k}{40}\boldsymbol{R}_u^7, \tag{27b}$$

$$\widetilde{\boldsymbol{v}}^{n+1} = \boldsymbol{v}^n + \frac{5179k}{57600}\boldsymbol{R}_v^1 + \frac{7571k}{16695}\boldsymbol{R}_v^3 + \frac{393k}{640}\boldsymbol{R}_v^4 - \frac{92097k}{339200}\boldsymbol{R}_v^5 + \frac{187k}{2100}\boldsymbol{R}_v^6 + \frac{k}{40}\boldsymbol{R}_v^7, \tag{27c}$$

is used for the error estimation. A predefined tolerance $\varepsilon$ is established such that the newly updated time step is computed based on the following condition (William and Saul, 1992)

$$k_{new} = \begin{cases} \eta k_{old}[\varepsilon/\max(e_u, e_v)]^{0.25}, & \max(e_u, e_v) < \varepsilon, \\ \eta k_{old}[\varepsilon/\max(e_u, e_v)]^{0.2}, & \max(e_u, e_v) \geq \varepsilon, \end{cases} \tag{28}$$

$\eta < 1$ and $\eta \approx 1$. Here, we choose $\eta = 0.9$. The new time step and solutions $\boldsymbol{u}^{n+1}$ and $\boldsymbol{v}^{n+1}$ are accepted as the optimal time step and numerical solutions, respectively, if $\max(e_u, e_v) < \varepsilon$. Else, we recompute our solutions with a smaller time step till the condition $\max(e_u, e_v) < \varepsilon$ is satisfied.

## 4. Numerical Results

### Example 1. Non-Dividend-Paying Options

We first consider non-dividend-paying options from the existing literature [24] with the following parameters

$$K = 100, \qquad r = 5\%, \qquad D = 0, \qquad \sigma = 20\%. \tag{29}$$

We chose the interval of $x \in [0,3]$. The focus in this example is to compute the convergence rate of our numerical method and verify that it is consistent with the theoretical convergence rate. Here, we computed the convergence rate with the fourth-order Runge-Kutta method, a constant time step of $k = 10^{-6}$ and varying step sizes $h = 0.1, 0.05, 0.025,$ and $0.0125$. The results were displayed in Table 1. From Table 1, we observed that the numerical convergence rate is in good agreement with the theoretical convergence rate.

**Table 1.** Errors and convergence rate in space with a third-order Robin boundary ($k = 10^{-6}, T = 0.25$).

| | Asset option | | Delta option | |
|---|---|---|---|---|
| $h$ | maximum error | convergence rate | maximum error | convergence rate |
| 0.1 | ~ | ~ | ~ | ~ |
| 0.05 | $9.105 \times 10^{-1}$ | ~ | $5.972 \times 10^{0}$ | ~ |
| 0.025 | $3.232 \times 10^{-2}$ | 4.777 | $3.290 \times 10^{-1}$ | 4.182 |
| 0.0125 | $2.075 \times 10^{-3}$ | 4.001 | $2.865 \times 10^{-2}$ | 3.521 |

| | Optimal exercise boundary | | |
|---|---|---|---|
| | values | maximum error | convergence rate |
| 0.1 | 87.744120699565300 | ~ | |
| 0.05 | 86.833647682786200 | $9.105 \times 10^{-1}$ | ~ |
| 0.025 | 86.803304297354500 | $3.034 \times 10^{-2}$ | 4.907 |
| 0.0125 | 86.805290298684800 | $1.986 \times 10^{-3}$ | 3.933 |

Even though implementation of the classical fourth-order time integration method could improve the accuracy of our numerical solution substantially, it could be tempting to assume that it is the reason why the fourth-order numerical convergence rate in space was recovered in our present method. We would rather love to point out that the recovery of the fourth-order convergence rate in space is due to our improvement and derivation in (13) and (14). To validate this claim, we implement the Crank-Nicholson scheme without further improvement in terms of strategic Rannacher time-stepping, mesh refinement, and smoothing of initial conditions. To this end, we first obtain an approximation of the optimal exercise boundary from the Crank-Nicholson scheme as follows:

$$f_b^n + \frac{k}{2}\frac{\partial f_b^n}{\partial \tau} = f_b^{n+1} - \frac{k}{2}\frac{\partial f_b^{n+1}}{\partial \tau}. \tag{30a}$$

Because the optimal exercise boundary is computed from the analytical approximation for the Crank-Nicholson scheme, we use the third-order analytical approximation as follows:

$$\frac{\partial f_b^n}{\partial \tau} = \left(\frac{-m_1^2 - \sqrt{m_2 - 4m_2 m_{0,3,n}}}{2g_0}\right) f_b^n(\tau) = \varpi^n f_b^n, \tag{30b}$$

$$f_b^n\left(1 + \frac{k}{2}\varpi^n\right) = f_b^{n+1}\left(1 - \frac{k}{2}\varpi^{n+1}\right), \qquad f_b^{n+1} = \frac{\left(1 + \frac{k}{2}\varpi^n\right)}{\left(1 - \frac{k}{2}\varpi^{n+1}\right)} f_b^n. \tag{30c}$$

$$m_2 = \frac{2L'(0,\tau)\gamma_2 \tilde{x}^3}{3\sigma^4}, \tag{30d}$$

$$m_1 = \left[\frac{4L'(0,\tau)\kappa}{3\sigma^4} - \frac{2Df_b(\tau)}{3\sigma^2 L'(0,\tau)}\right]\gamma_2 \tilde{x}^3 - \frac{2L'(0,\tau)\gamma_1 \tilde{x}^2}{3\sigma^2}, \tag{30e}$$

$$\begin{aligned} m_{0,3} = &\left[\frac{2L'(0,\tau)\kappa^2}{3\sigma^4} + \frac{Df_b(\tau)\kappa}{6\sigma^4 L'(0,\tau)} - \frac{\left(Df_b(\tau)\right)^2}{12\sigma^4[L'(0,\tau)]^3} + \frac{rL'(0,\tau)}{2\sigma^2} - \frac{Df_b(\tau)}{4\sigma^2 L'(0,\tau)}\right]\gamma_2 \tilde{x}^3 \\ &- \left[\frac{2L'(0,\tau)}{3\sigma^2}\kappa + \frac{Df_b(\tau)}{3\sigma^2 L'(0,\tau)}\right]\gamma_1 \tilde{x}^2 + \gamma_0 \tilde{x} L'(0,\tau) - M_3(\tilde{x},\tau), \end{aligned} \tag{30f}$$

$$M_3(\tilde{x},\tau) = \alpha_1 L(\tilde{x},\tau) + \alpha_2 L(2\tilde{x},\tau) + \alpha_3 L(3\tilde{x},\tau). \tag{30g}$$

$$\alpha_1 = 81, \qquad \alpha_2 = -\frac{81}{8}, \qquad \alpha_4 = 1; \tag{30h}$$

$$\gamma_0 = \frac{255}{4}, \qquad \gamma_1 = \frac{99}{4}, \qquad \gamma_2 = \frac{9}{2}. \tag{30i}$$

From (23), we computed the left-hand boundary values of (3a) and (3b) and discretized the interior nodes in time with the Crank-Nicholson scheme. We then computed the convergence rate of the optimal exercise boundary, asset option, and option sensitivity which we listed in Table 2. From Table 2, one can easily observe that the convergence rate in space with the Crank-Nicholson discretization in time is in close agreement with the theoretical convergence rate. The major advantage of implementing 5(4) Runge-Kutta embedded pairs is to employ a fifth-order time integration method and allow optimal time step selection in each time level. The latter advantage could further be very useful in detecting unknown locations with large variations in a system due to discontinuity, oscillation, etc.

**Table 2.** Errors and convergence rate in space with third-order Robin boundary and Crank-Nicholson scheme. ($k = 10^{-6}, T = 0.5$).

| | Asset option | | Delta option | |
|---|---|---|---|---|
| $h$ | maximum error | convergence rate | maximum error | convergence rate |
| 0.1 | ~ | ~ | ~ | ~ |
| 0.05 | $7.452 \times 10^{-1}$ | ~ | $1.173 \times 10^{0}$ | ~ |
| 0.025 | $4.650 \times 10^{-2}$ | 4.002 | $2.363 \times 10^{-1}$ | 2.311 |
| 0.0125 | $3.300 \times 10^{-3}$ | 3.817 | $1.362 \times 10^{-2}$ | 4.117 |
| 0.00625 | $2.171 \times 10^{-4}$ | 3.926 | $8.179 \times 10^{-4}$ | 4.057 |

| | Optimal exercise boundary | | |
|---|---|---|---|
| | values | maximum error | convergence rate |
| 0.1 | 84.6168895718962 | ~ | |
| 0.05 | 83.8717223863237 | $7.452 \times 10^{-1}$ | ~ |
| 0.025 | 83.9162106246566 | $4.449 \times 10^{-2}$ | 4.002 |
| 0.0125 | 83.9193524617783 | $3.142 \times 10^{-3}$ | 3.824 |
| 0.00625 | 83.9195594221456 | $2.070 \times 10^{-4}$ | 3.924 |

### Example 2. Dividend-Paying Options

For the dividend-paying American options, we consider two examples from the work of Kim et al. [24] and Tangman et al. [38] with the following parameters:

$$K = 100, \quad T = 0.5, \quad r = 5\%, \quad D = 3\ \%, \quad \sigma = 20\%, \tag{31a}$$

$$K = 100, \quad T = 0.5, \quad r = 7\%, \quad D = 3\%, \quad \sigma = 40\%. \tag{31b}$$

It is well known that the performance and accuracy of the adaptive Runge-Kutta methods depend on the type of model under investigation. Not much has been done on implementing these embedded pairs in the option pricing problem. For our present model with a discontinuity at the left boundary and time-dependent coefficient, it will be ideal to extensively compare the existing 5(4) Runge-Kutta embedded pair to validate the one that performs optimally. The first objective in this example is to compare the performance of our present method with several 5(4) embedded pairs. The second objective is to compare our numerical results of the asset option, option sensitivity, and optimal exercise boundary with the existing methods. To this end, we label the Runge-Kutta 5(4) pairs as follows:

- Dormand and Prince 5(4) embedded pair (RK-DR)
- Modified 5(4) Runge-Kutta-Fehlberg method based on Cash and Karp coefficients (RK-CK)
- Simon and Tsitorous 5(4) embedded pair (RK-ST)
- Bogacki and Shampine 5(4) embedded pair (RK-BS)
- Papakostas and Papageorgiou 5(4) embedded pair (RK-NEW)

With (24a), we compared the performance and accuracy of several well-known 5(4) Runge-Kutta pairs. These pairs are fifth-order accurate in time. We intentionally use a large tolerance ($\varepsilon = 10^{-3}, \varepsilon = 10^{-4}$, or $\varepsilon = 10^{-5}$) in most cases in this example. This is because we intend to investigate the performance of these embedded pairs with varying large tolerances and step sizes. We also compared the numerical results of the asset option from our present method with the method of Wu and Kwok [42], Kim et al. [24], and Muthuraman [29] which we label as WK, KIM, and MBM, respectively. We chose the method of Cox et al. [12] as the benchmark result. Furthermore, we used a very small step size to compute the optimal exercise boundary. Using the former as a benchmark, we then compared the results of the optimal exercise boundary obtained with these pairs when $\varepsilon$ and $h$ were varied. We also compared the total CPU time required to obtain the numerical results of the optimal exercise boundary with these pairs. The results were listed in Tables 3-5 and displayed in Fig. 1.

With a tolerance of $\varepsilon = 10^{-5}$, we observed from Table 3 that we have a more accurate numerical solution when $h = 0.025$ with RK-DP and RK-BS. Furthermore, with $h = 0.01$ and $10^{-4}$, the optimal exercise boundary obtained with RK-DP and RK-BS is very close to the benchmark result and requires less total CPU time and average time step when compared to RK-ST and RK-CK. Hence RK-DP and RK-BS present more superior advantages in terms of accuracy and computational cost. It is worth mentioning that we obtained numerical divergence with RK-NEW.

With (24b), we used very large step sizes to obtain the numerical solution of the delta option and compared the result with the methods of Brennan and Schwartz [5], Han and Wu [19], Ikonen and Toivanen [22], and Forsyth and Vetzal [16] which we label as BS, HW, OS, and PENALTY, respectively. The Binomial method of Leisen and Reimer [26] is used as the benchmark result. The results were listed in Table 6.

With $h = 0.075$ and $\varepsilon = 10^{-5}$, the results we obtained from our pairs are very close to the benchmark result. This is an indication of the superiority of our approach in terms of achieving accuracy with very coarse grids. One of the reasons for the highly accurate result of the delta option is because we computed the latter simultaneously with the asset using a fourth-order compact operator based on our formulation in (4). In terms of total CPU time in seconds, RK-DP and RK-BS are superior to RK-CK AND RK-ST. We also plotted the time step against the time level and displayed the result in Fig. 2. The common characteristic in the plots is that a very small time step is required very close to the fixed free boundary. This is because of the piecewise nature of the asset option and the discontinuity in the delta option at the payoff. Hence small time would be required in that region. This is a well-known behavior of RK-embedded pairs. In Fig. 2, the horizontal red line represents the average optimal time step. We further observed that this value is smaller in RK-DP and RK-BS when compared with RK-ST and RK-CK.

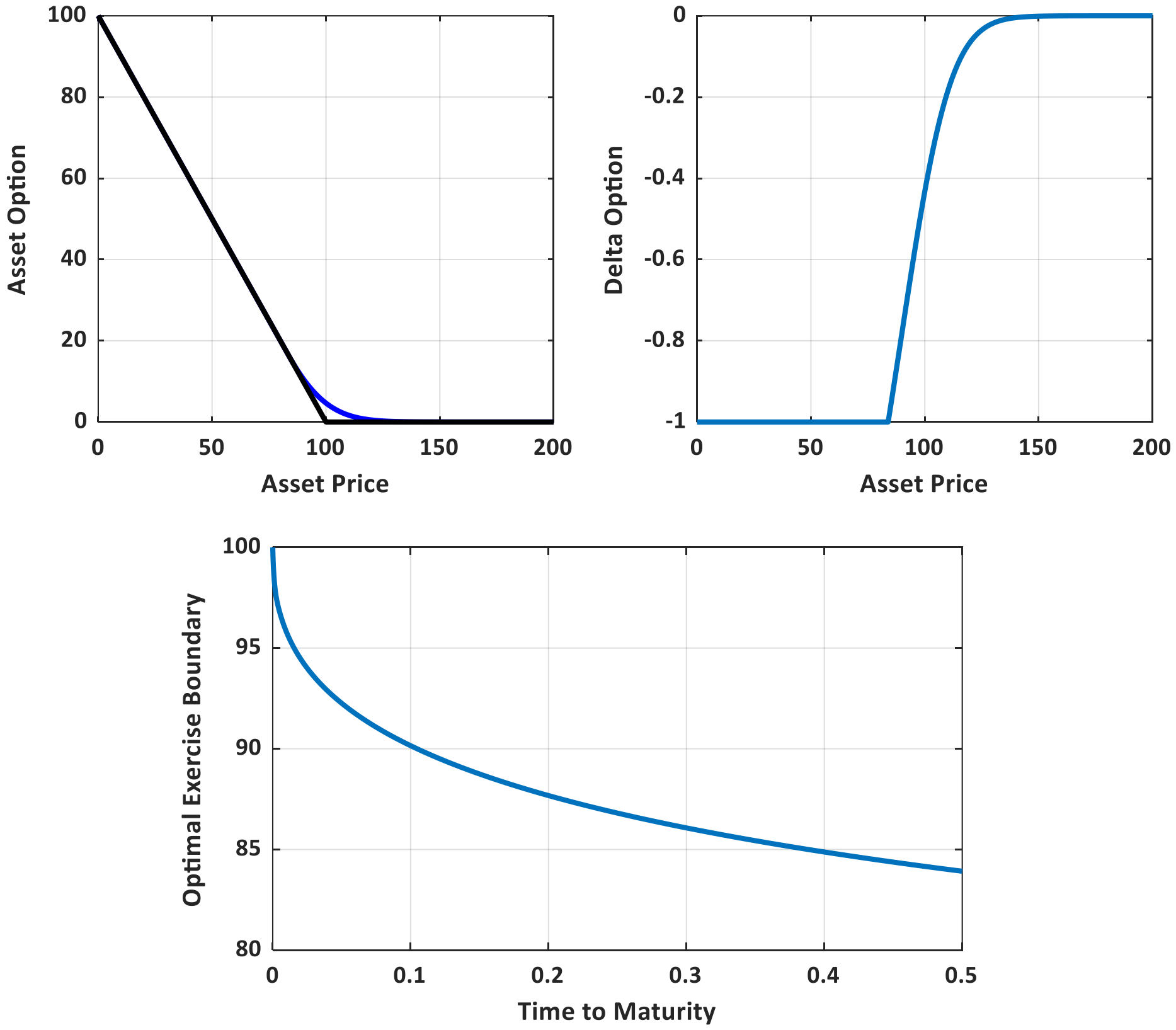

Fig. 1a. Plots of the option values and optimal exercise boundary in (24a) without dividend ($\tau = T$).

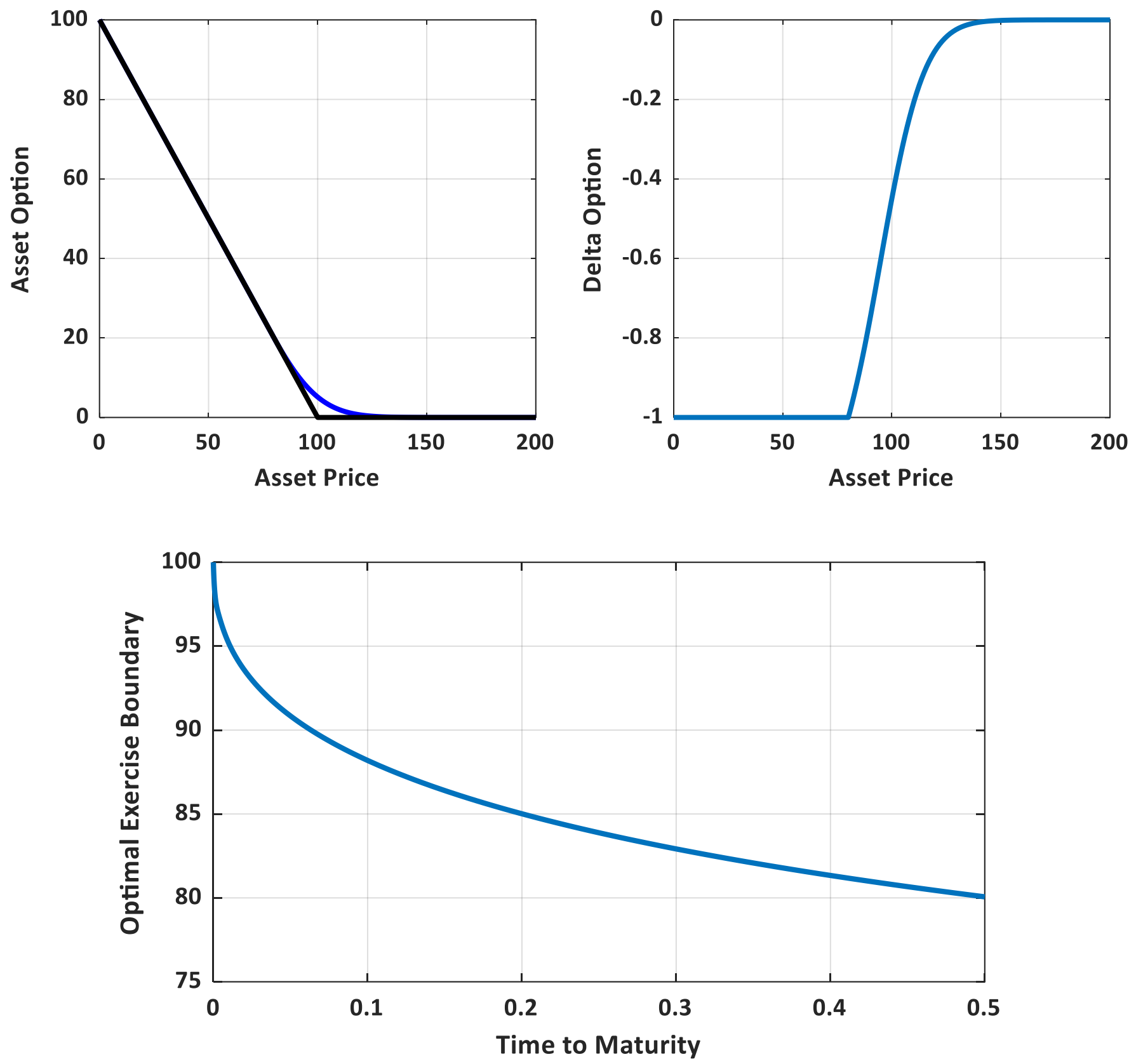

Fig. 1b. Plots of the option values and optimal exercise boundary in (24a) with dividend ($\tau = T$).

**Table 3.** Comparison of the asset option with (24a) and the embedded 5(4) RK pairs ($\varepsilon = 10^{-5}, k = h$).

| S | True Value | WK | MBM | KIM |
|---|---|---|---|---|
| 80 | 20.0000 | 20.0000 | 20.0000 | 20.0000 |
| 90 | 11.1551 | 11.1513 | 11.1526 | 11.1544 |
| 100 | 5.1496 | 5.1435 | 5.1444 | 5.1496 |
| 110 | 1.9491 | 1.9461 | 1.9455 | 1.9509 |
| 120 | 0.6132 | 0.6113 | 0.6155 | 0.6153 |

| | RK-DP | | | RK-CK | | |
|---|---|---|---|---|---|---|
| | $h = 0.025$ | 0.0125 | 0.01 | 0.025 | 0.0125 | 0.01 |
| 80 | 20.0000 | 20.0000 | 20.0000 | 20.0000 | 20.0000 | 20.0000 |
| 90 | 11.1550 | 11.1548 | 11.1548 | 11.1550 | 11.1547 | 11.1547 |
| 100 | 5.1494 | 5.1494 | 5.1494 | 5.1489 | 5.1489 | 5.1490 |
| 110 | 1.9487 | 1.9487 | 1.9488 | 1.9477 | 1.9480 | 1.9481 |
| 120 | 0.6130 | 0.6130 | 0.6131 | 0.6121 | 0.6124 | 0.6125 |

| | RK-ST | | | RK-BS | | | RK-NEW | |
|---|---|---|---|---|---|---|---|---|
| | $h = 0.025$ | 0.0125 | 0.01 | 0.025 | 0.0125 | 0.01 | 0.0125 | 0.01 |
| 80 | 20.0000 | 20.0000 | 20.0000 | 20.0000 | 20.0000 | 20.0000 | diverge | diverge |
| 90 | 10.1550 | 10.1547 | 10.1547 | 10.1550 | 10.1548 | 10.1548 | diverge | diverge |
| 100 | 5.1488 | 5.1489 | 5.1490 | 5.1494 | 5.1494 | 5.1494 | diverge | diverge |
| 110 | 1.9474 | 1.9479 | 1.9481 | 1.9487 | 1.9487 | 1.9488 | diverge | diverge |
| 120 | 0.6118 | 0.6123 | 0.6125 | 0.6130 | 0.6130 | 0.6131 | diverge | diverge |

**Table 4.** Values of the optimal exercise boundary with RK-DP and very small step sizes.

| $h$ | $0.01$ | $0.005$ | $0.0025$ |
|---|---|---|---|
| $f_b(\tau)$ | 80.06279138725 | 80.06250056775 | 80.06233787425 |

**Table 5a.** Performance of the embedded pairs on fixed step size ($h = 0.01, k = h$) and varying tolerance.

| | RK-DP | | | RK-CK | | |
|---|---|---|---|---|---|---|
| $\varepsilon$ | $10^{-3}$ | $10^{-4}$ | $10^{-5}$ | $10^{-3}$ | $10^{-4}$ | $10^{-5}$ |
| Total CPU time(s) | 11.81 | 12.33 | 13.38 | 23.73 | 64.31 | 187.79 |
| $f_b(\tau)$ | 80.0621 | 83.0628 | 83.0628 | 80.0674 | 80.0637 | 80.0631 |
| Min. optimal time step | 2.72e-5 | 7.57e-6 | 2.25e-6 | 4.06e-6 | 1.26e-6 | 3.95e-7 |
| Ave. optimal time step | 1.85e-3 | 1.79e-3 | 1.69e-3 | 7.86e-4 | 2.77e-4 | 8.99e-5 |
| Max. optimal time step | 2.62e-3 | 2.13e-3 | 2.14e-3 | 2.19e-3 | 7.82e-4 | 2.64e-4 |
| | RK-ST | | | RK-BS | | |
| $\varepsilon$ | $10^{-3}$ | $10^{-4}$ | $10^{-5}$ | $10^{-3}$ | $10^{-4}$ | $10^{-5}$ |
| Total CPU time(s) | 12.92 | 20.01 | 38.24 | 17.84 | 18.57 | 19.48 |
| $f_b(\tau)$ | 80.0681 | 80.0645 | 80.0630 | 80.0658 | 80.0628 | 80.0628 |
| Min. optimal time step | 3.45e-5 | 6.95e-6 | 4.88e-6 | 7.78e-5 | 2.41e-5 | 7.23e-6 |
| Ave. optimal time step | 1.61e-3 | 9.73e-4 | 5.29e-4 | 2.36e-3 | 2.31e-3 | 2.30e-3 |
| Max. optimal time step | 2.21e-2 | 6.84e-3 | 1.43e-3 | 4.42e-3 | 5.46e-2 | 7.37e-3 |

**Table 5b.** Performance of the embedded pairs on fixed tolerance and varying step size ($\varepsilon = 10^{-5}, k = h$).

| | RK-DP | | | RK-CK | | |
|---|---|---|---|---|---|---|
| $h$ | $0.025$ | $0.0125$ | $0.01$ | $0.025$ | $0.0125$ | $0.01$ |
| Total CPU time(s) | 1.21 | 6.15 | 13.38 | 27.13 | 105.25 | 187.79 |
| $f_b(\tau)$ | 80.0611 | 80.0628 | 80.0628 | 80.0604 | 80.0629 | 80.0631 |
| Min. optimal time step | 1.93e-5 | 3.79e-6 | 2.25e-6 | 2.54e-6 | 6.58e-7 | 3.95e-7 |
| Ave. optimal time step | 6.41e-3 | 2.45e-3 | 1.69e-3 | 1.59e-4 | 1.04e-5 | 8.99e-5 |
| Max. optimal time step | 1.25e-2 | 3.36e-3 | 2.14e-3 | 6.23e-4 | 3.29e-4 | 2.64e-4 |
| | RK-ST | | | RK-BS | | |
| $h$ | $0.025$ | $0.0125$ | $0.01$ | $0.025$ | $0.0125$ | $0.01$ |
| Total CPU time(s) | 5.03 | 20.36 | 38.24 | 1.22 | 7.89 | 19.48 |
| $f_b(\tau)$ | 80.0603 | 80.0628 | 80.0630 | 80.0611 | 80.0629 | 80.0628 |
| Min. optimal time step | 4.41e-5 | 8.08e-6 | 4.88e-6 | 6.23e-8 | 1.22e-5 | 7.23e-6 |
| Ave. optimal time step | 1.00e-3 | 6.36e-4 | 5.29e-4 | 9.43e-3 | 3.40e-3 | 2.30e-3 |
| Max. optimal time step | 4.22e-3 | 2.00e-3 | 1.43e-3 | 3.07e-2 | 1.10e-2 | 7.37e-3 |

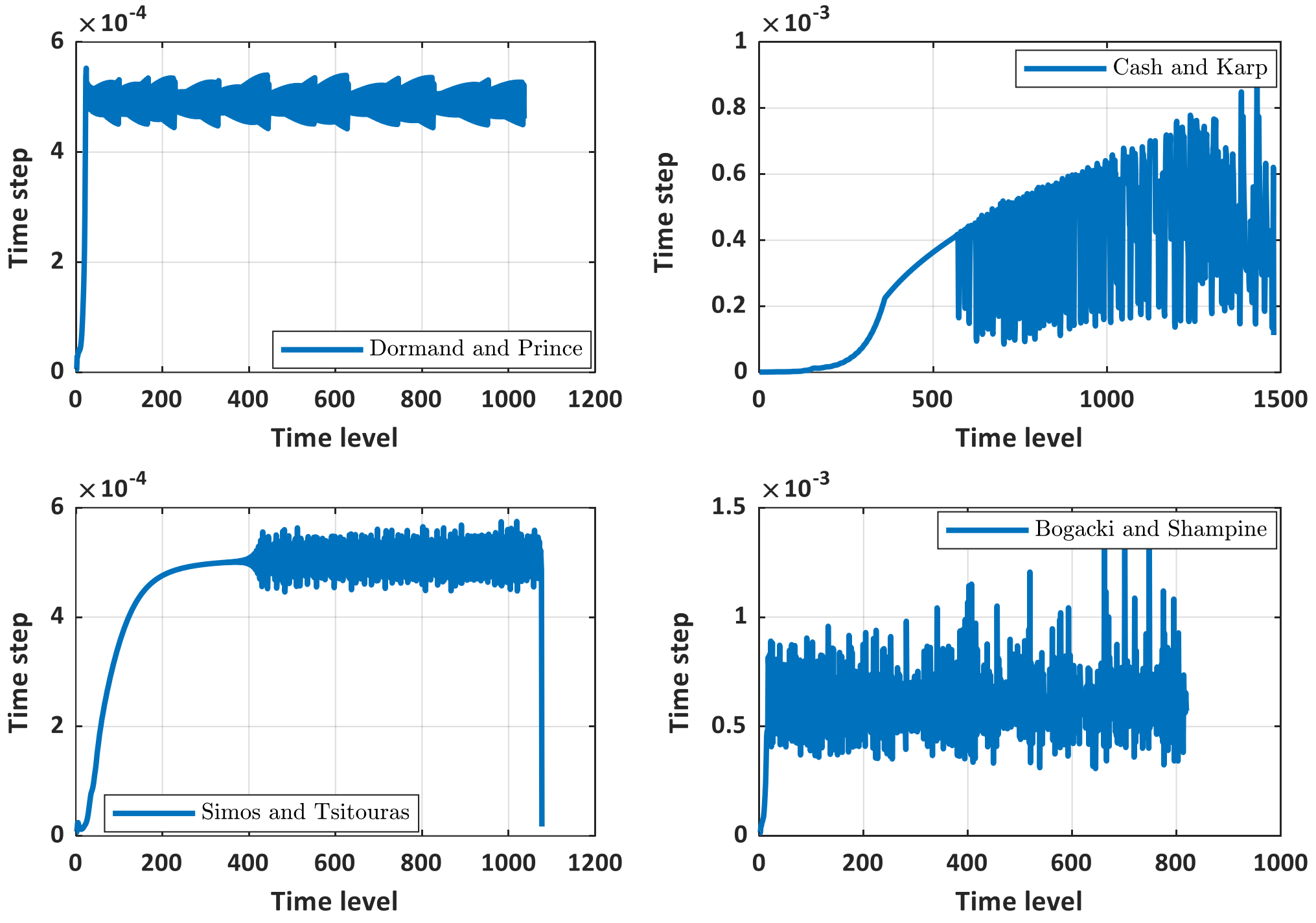

Fig. 2a. Optimal time step in each time level with $\varepsilon = 10^{-3}$, $h = 0.01$ and $k = h$.

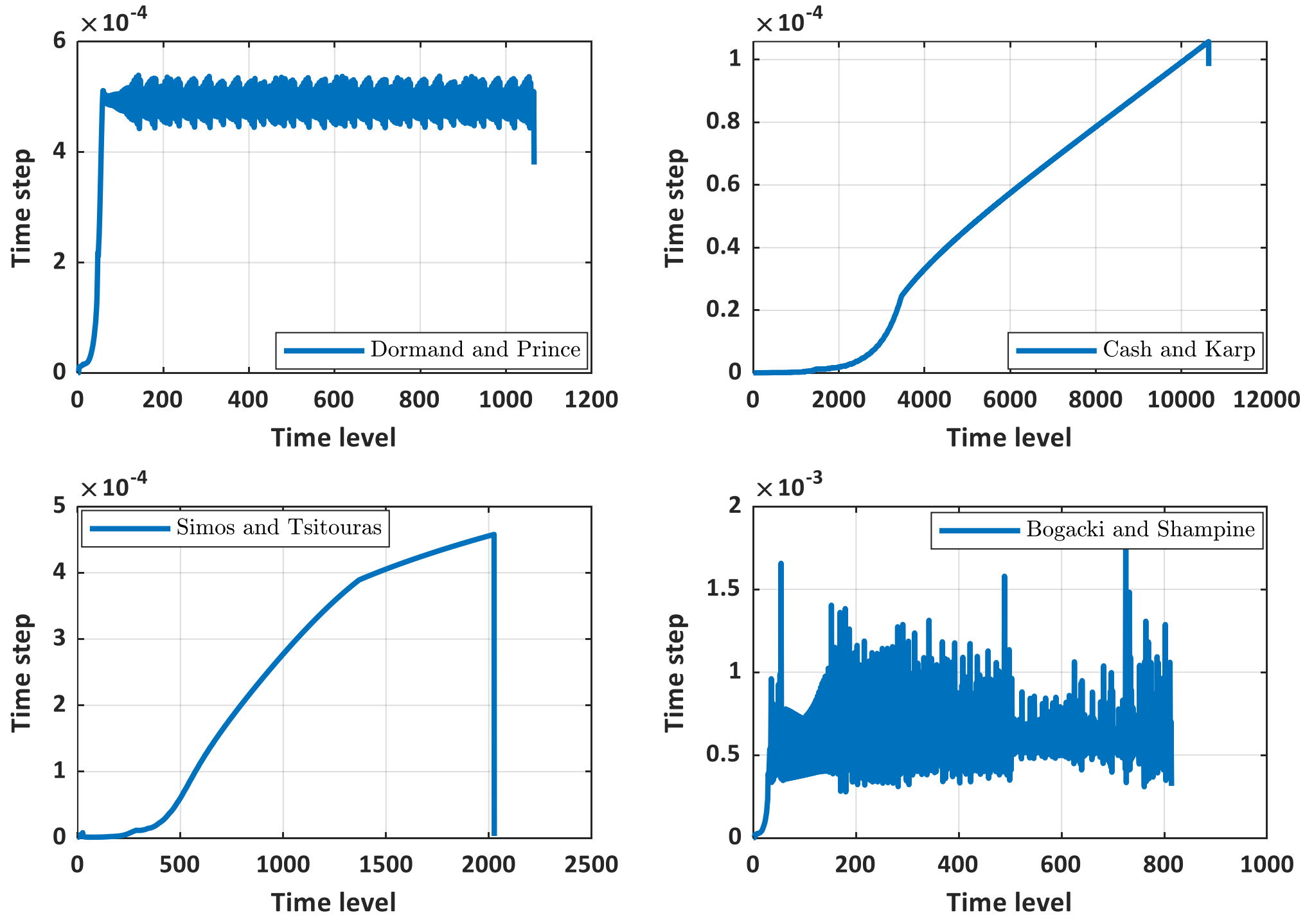

Fig. 2b. Optimal time step in each time level with $\varepsilon = 10^{-5}$, $h = 0.01$ and $k = h$.

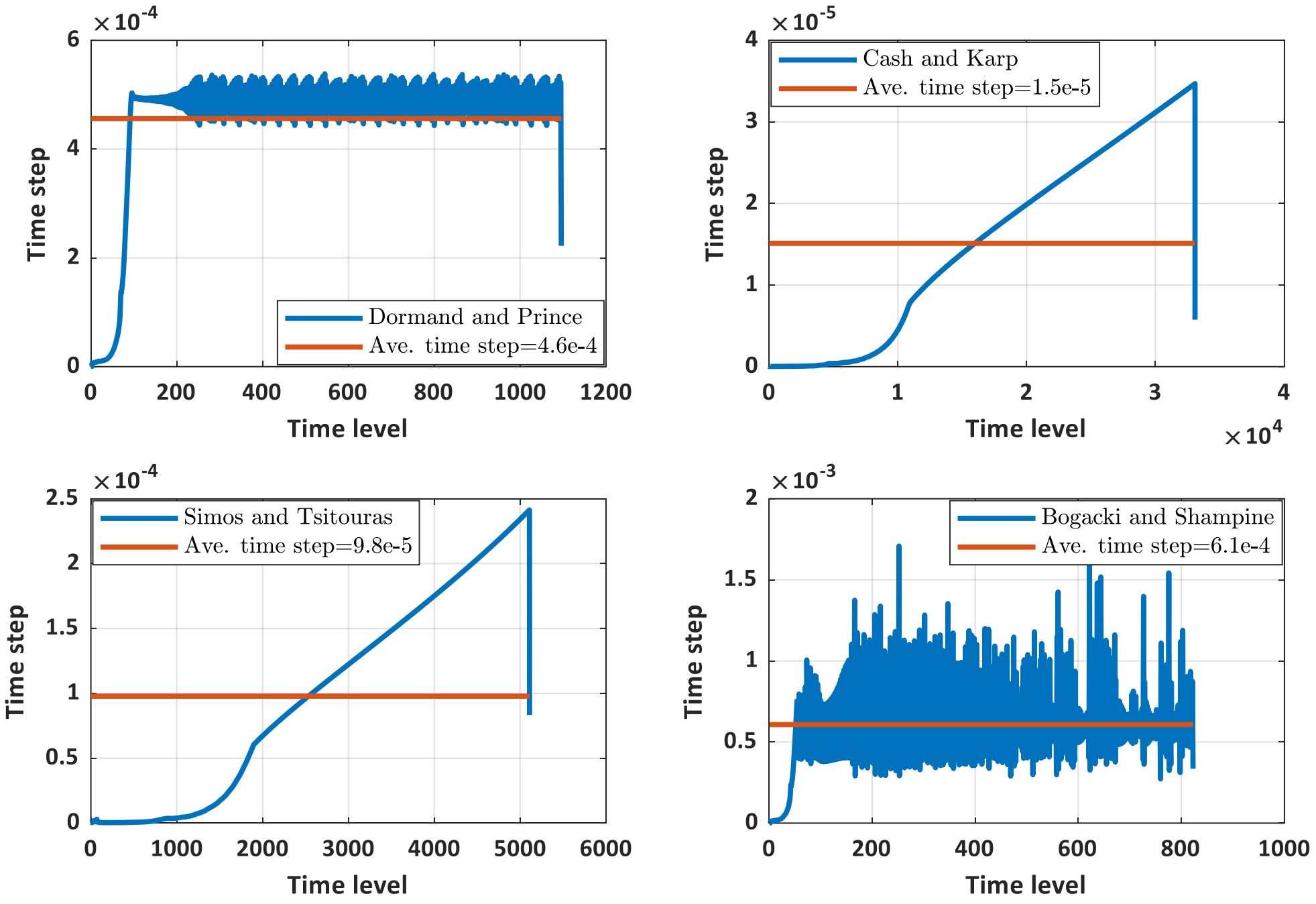

Fig. 2c. Optimal time step in each time level and average time step with $\varepsilon = 10^{-6}$, $h = 0.01$ and $k = h$.

**Table 6.** Comparison of the delta option with (24b) and the embedded 5(4) RK pairs ($\varepsilon = 10^{-5}, k = h$).

| S | True Value | BS2 | HW | OS | PENALTY | |
|---|---|---|---|---|---|---|
| 80 | -0.7501 | -0.7501 | -0.7489 | -0.7501 | -0.7502 | |
| 90 | -0.5791 | -0.5791 | -0.5791 | -0.5791 | -0.5791 | |
| 100 | -0.4229 | -0.4229 | -0.4222 | -0.4230 | -0.4229 | |
| 110 | -0.2943 | -0.2943 | -0.2938 | -0.2943 | -0.2943 | |
| 120 | -0.1968 | -0.1968 | -0.1965 | -0.1968 | -0.1968 | |
| | | RK-DP | | | RK-CK | |
| | $h = 0.075$ | 0.05 | 0.03 | 0.075 | 0.05 | 0.03 |
| 80 | -0.7500 | -0.7501 | -0.7501 | -0.7501 | -0.7501 | -0.7502 |
| 90 | -0.5792 | -0.5791 | -0.5791 | -0.5793 | -0.5792 | -0.5791 |
| 100 | -0.4231 | -0.4230 | -0.4229 | -0.4231 | -0.4230 | -0.4229 |
| 110 | -0.2944 | -0.2943 | -0.2943 | -0.2944 | -0.2943 | -0.2943 |
| 120 | -0.1968 | -0.1968 | -0.1968 | -0.1968 | -0.1968 | -0.1968 |
| Total CPU time(s) | 0.50 | 0.79 | 1.28 | 6.26 | 9.67 | 22.25 |
| | | RK-ST | | | RK-BS | |
| | $h = 0.075$ | 0.05 | 0.03 | 0.075 | 0.05 | 0.03 |
| 80 | -0.7502 | -0.7502 | -0.7502 | -0.7500 | -0.7501 | -0.7502 |
| 90 | -0.5792 | -0.5792 | -0.5792 | -0.5793 | -0.5791 | -0.5791 |
| 100 | -0.4230 | -0.4230 | -0.4230 | -0.4231 | -0.4230 | -0.4229 |
| 110 | -0.2943 | -0.2943 | -0.2943 | -0.2944 | -0.2943 | -0.2943 |
| 120 | -0.1968 | -0.1968 | -0.1968 | -0.1968 | -0.1968 | -0.1968 |
| Total CPU time(s) | 1.12 | 2.37 | 5.45 | 0.59 | 0.76 | 1.88 |

## 5. Conclusion

We have proposed an efficient and high-order numerical method for pricing free boundary non-dividend and dividend-paying options. The major challenge in solving the free boundary option pricing model is to recover the high order convergence rate. It has been shown in previous literature that even with an efficient fourth-order numerical scheme, only second-order convergence can be expected. To recover the high order convergence rate from our present method, we derive a high order analytical approximation from the left boundary and obtained a precise value of the time-dependent coefficient that introduces nonlinearity in the model. Furthermore, we introduce a special boundary treatment that enables us to compute the optimal exercise boundary without further discretization in time. With the implementation of a compact finite difference scheme, we then carried further investigation on the performance of our present method with several 5(4) RK-embedded pairs. This is by no means an exhaustive investigation; however, it gives us an insight into how some of these pairs can significantly improve both the accuracy and computational cost of our numerical solutions. One of the reasons for this investigation is because some of these pairs are model-dependent. From the numerical experiment, we observe that RK-DP and RK-BS perform significantly better than RK-TS and RK-CK. Moreover, we achieve numerical divergence with RK-NEW. We compute the convergence rate in space of our present method and validate that the numerical convergence rate is in good agreement with the theoretical convergence rate. By further comparing with other existing methods using the example in (24a) and (24b), we confirm that our present method provides more accurate numerical solutions with very coarse grids.

## Acknowledgment

This is a preprint of an article published in Japan Journal of Industrial and Applied Mathematics (JJIAM). The final authenticated version is available online at: https://doi.org/10.1007/s13160-022-00507-0.

## Funding

No funding was received from any resource.

## Availability of Data and Material (Data Transparency)

The data could be shared at a reasonable request.

## Conflict of interest

The author declares that he has no conflict of interest.